\documentclass[a4paper, 10 pt]{article}

\usepackage[top = 1 in, bottom = 1 in, left = 1 in, right = 1 in]{geometry}
\usepackage{amsmath, amssymb, amsthm}
\usepackage{hyperref, multirow}
\usepackage{pgfplots, pgfplotstable}
\usepgfplotslibrary{external, statistics}
\usepackage{tikz}
\usetikzlibrary{shapes, arrows, positioning, plotmarks, patterns}

\hypersetup{
colorlinks = true,
linkcolor = black,
citecolor = black,
filecolor = black,
urlcolor = black,
bookmarks = true,
pdftoolbar = true,
pdfmenubar = true,
pdfstartview = {FitH},
pdftitle = {Traffic Delay Reduction at Highway Diverges Using an Advance Warning System Based on a Probabilistic Prediction Model},
pdfauthor = {Goodarz Mehr},
pdfsubject = {Diverge Advisory},
}

\begin{document}
	
	\title{Traffic Delay Reduction at Highway Diverges Using an Advance Warning System Based on a Probabilistic Prediction Model}
	
	\author{Goodarz Mehr\thanks{Department of Mechanical Engineering, Virginia Tech, Blacksburg, VA 24061, USA. Email: goodarzm@vt.edu}, Azim Eskandarian\thanks{Nicholas and Rebecca Des Champs Mechanical Engineering Department Chair, Virginia Tech, Blacksburg, VA 24061, USA.}}

	\maketitle

	\begin{abstract}
		
		This paper presents an on-board advance warning system for vehicles based on a probabilistic prediction model that advises them on when to change lanes to reach a highway diverge on time. The system is based on a model that estimates the probability of reaching a goal state on the road using one or multiple lane changes. This estimate is based on several traffic-related parameters such as the distribution of inter-vehicle headway distances as well as driver-related parameters like lane change duration. For an upcoming diverge, the advance warning system uses the model to continuously calculate the probability of reaching it and advise the driver to change lanes when the probability dips below a certain threshold. To evaluate the performance of the proposed system in reducing traffic delay at highway diverges, it was used on a segment of a four-lane highway to advise vehicles taking an off-ramp on when to change lanes. Results show that using the proposed system reduces average delay up to 6\% and maximum delay up to 16\%, depending on traffic flow and the ratio of vehicles taking the off-ramp.
	
	\end{abstract}
	
	\section{Introduction} \label{Section1}
	
	Lane changes are essential to highway driving. These maneuvers are the primary method of navigation in highways and are influenced by multiple factors, including driving behavior, urgency of changing lanes, and the state of nearby vehicles \cite{Brackstone}. For a successful maneuver the driver (or autonomous vehicle) has to identify an acceptable gap in the target lane, adjust speed and maintain correct position relative to nearby vehicles, and navigate to the target lane while avoiding collision with other vehicles \cite{Kesting}. Because of this, unsafe driving behavior or a small mistake at any step can result in an accident. In the U.S. between four to ten percent of all reported motor vehicle crashes are due to unsafe lane change behavior. Apart from the fatalities, these crashes incur an economic and productivity loss by delaying hundreds of vehicles \cite{Sen, Lisheng, Dijck}. This can be mitigated if vehicles obtain accurate and timely information to advise them on when to change lanes. \par
	
	Lane changes are classified as either discretionary or mandatory \cite{Zhang}. Discretionary lane changes are often performed to move to a lane with a higher speed and overtake slow traffic. In contrast, mandatory lane changes are required to follow a planned path, for example to reach a highway diverge. Compared to discretionary lane changes, mandatory lane changes can have a disruptive impact on traffic. Mandatory lane changes can cause traffic oscillation \cite{Sarvi}, traffic breakdown \cite{Lv}, capacity drops \cite{Cassidy}, and deteriorate traffic safety \cite{Ahammed, Li}. \par
	
	To manage traffic upstream of a diverge, past studies have identified advance warning systems as a way to decrease the number of unsafe lane changes and reduce traffic delay \cite{Gong, Hang, He, Yun}. Gong et al. determined the optimal location of advance warning in a two-lane highway and divided its downstream area into two zones; a green zone whose traffic ensures lane changes without deceleration and a yellow zone whose traffic leads to rushed lane changes and speed deceleration \cite{Gong}. Optimizing for the advance warning point in a numerical simulation, they observed that it minimized the corresponding traffic delay and mitigated capacity drop and traffic oscillation. Hang et al. used a driving simulator to study the effects of advance warning location in work zone areas on lane changing behavior and found that it had a strong impact on drivers' perception of the imminent situation \cite{Hang}. Similarly, He et al. and Yun et al. found that providing timely warning can reduce delay in moderate and congested traffic \cite{He, Yun}. \par
	
	While these studies made great strides by introducing methods to reduce traffic delay at diverges, the proposed methods have some limitations. For one, they are static, in the sense that they determine only one location where drivers are warned of an upcoming diverge, either by a road sign or an in-vehicle signal. The problem is that a changing traffic flow will require new calculations to determine the new warning point. More importantly, these methods only apply to highways with two lanes and do not generalize to those with more. \par

	To address these problems, we propose a new advance warning system for vehicles in a highway based on a probabilistic prediction model that advises them on when to change lanes to reach a highway diverge on time \cite{Mehr}. The model estimates the probability of reaching a goal state on the road using one or multiple lane changes. This estimate is based on several traffic-related parameters such as average vehicle velocity and the distribution of inter-vehicle headway distances on each lane as well as driver-related parameters like lane change duration. For an upcoming diverge, the advance warning system uses the model to continuously calculate the probability of reaching it and advises the driver to change lanes when the probability dips below a certain threshold. In other words, the system warns the driver when the probability becomes low enough that the driver may miss the exit unless he/she acts soon. The model used has real-time performance and can be applied to highways with any number of lanes, addressing the limitations of previous methods. \par
	
	The remainder of this paper is structured as follows. \autoref{Section2} presents the methodology, including a brief overview of the probabilistic model and the simulation setup used to evaluate the performance of the advance warning system. \autoref{Section3} presents our results and a discussion of the effects of the proposed system on traffic flow. Finally, \autoref{Section4} concludes the findings of this paper.
	
	\section{Methodology} \label{Section2}
	
	The advance warning system proposed in this paper is based on a model that predicts the probability of reaching a near-term goal state using one or multiple lane changes \cite{Mehr}. Using that model, the system advises the vehicle on when to change lanes. \autoref{Section2.1} reviews the probability model, while \autoref{Section2.2} describes the advance warning system based on that model and the simulation setup used to evaluate the effectiveness of the advance warning system in reducing traffic delay at highway diverges.
	
	\subsection{Probability model} \label{Section2.1}
	
	The model introduced in \cite{Mehr} estimates the probability of reaching a near-term goal state using one or multiple lane changes. While a brief overview of the model is provided here, detailed derivation and validation of the model can be found in \cite{Mehr}. \par
	
\begin{figure}[t!]
		\centering
		\includegraphics[width = \textwidth]{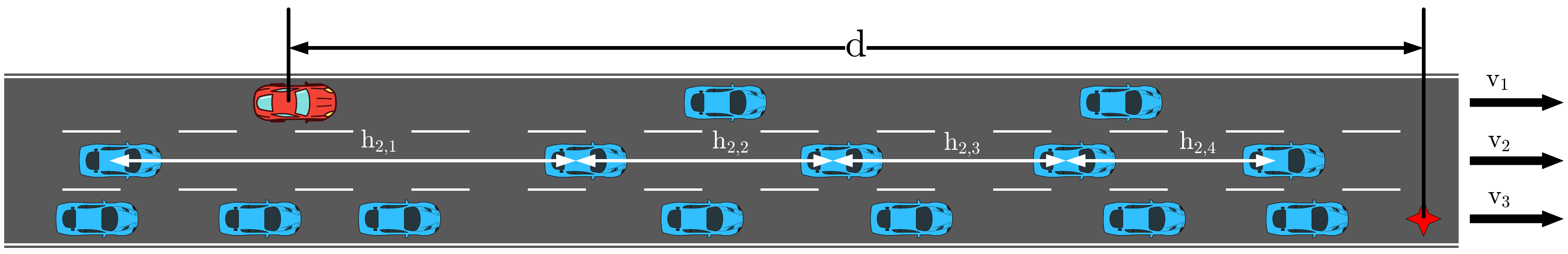}
		\caption{Notations used in this paper for a road segment with three lanes. The red car is the ego vehicle and the red star shows the goal state.} \label{Figure1}
	\end{figure}

	Without loss of generality, consider a highway with $n$ lanes, numbered by 1 to $n$ from left to right. Assume that the ego vehicle wants to reach a position on lane $n$ a distance $d$ ahead of its current position on lane 1. Denoting the success probability of doing so by $P(S)$, the model estimates this probability by making a few assumptions. First, the model assumes that the velocity of all vehicles on lane $i$, $1 \le i \le n$, is equal to $v_{i}$, where $v_{i}$ is the average velocity of all vehicles on that lane over a period of time. Second, the model assumes that inter-vehicle headway distances (front bumper to front bumper) on lane $i$ are independent identically distributed (i.i.d) random variables from a common log-normal distribution defined by parameters $\mu_{i}$ and $\sigma_{i}$ \cite{Mei}. Finally, the model assumes that the ego vehicle follows a Gipps gap acceptance model when changing lanes \cite{Gipps}. That is, if the ego vehicle is on lane $i - 1$, it only changes lanes if the gap between its leading and trailing vehicles on the adjacent lane $i$ is no smaller than a minimum acceptable (ciritical) gap $g_{i}$. Such a lane change takes $t_{i}$ seconds to complete. For better visualization, some of these assumptions are shown in \autoref{Figure1}.
	
	The model estimates the success probability $P(S)$ based on the parameters defined previously. In other words, for the case described above $P(S) = f_{n}(d, v_{1 : n}, \mu_{2 : n}, \sigma_{2 : n}, g_{2 : n}, t_{2 : n})$, where $w_{l : m}$ means $w_{l}, w_{l + 1}, \ldots, w_{m}$ for any parameter $w$ and indexes $m \ge l$. $P(S)$ is estimated recursively, with $n = 2$ as the base case. For $n = 2$, $P(S)$ is obtained from a look-up table of values calculated by Monte Carlo simulations of a normalized problem, since a closed-from expression for the probability does not exist. For $n > 2$, $P(S)$ is obtained recursively from
	\begin{equation} \label{Equation1}
		\begin{split}
			&f_{n}(d, v_{1 : n}, \mu_{2 : n}, \sigma_{2 : n}, g_{2 : n}, t_{2 : n})\\
			&= \int_{0} ^ {d} f_{2}(d - x, v_{n - 1 : n}, \mu_{n}, \sigma_{n}, g_{n}, t_{n})\frac{\partial}{\partial x} f_{n - 1}(x, v_{1 : n - 1}, \mu_{2 : n - 1}, \sigma_{2 : n - 1}, g_{2 : n - 1}, t_{2 : n - 1})\mathrm{d}x \\
			&= \frac{\partial}{\partial x}\int_{0} ^ {d} f_{2}(d - x, v_{n - 1 : n}, \mu_{n}, \sigma_{n}, g_{n}, t_{n})f_{n - 1}(x, v_{1 : n - 1}, \mu_{2 : n - 1}, \sigma_{2 : n - 1}, g_{2 : n - 1}, t_{2 : n - 1})\mathrm{d}x,
		\end{split}
	\end{equation}
	which is based on the law of total probability \cite{Leon}. Extensive traffic simulations in VISSIM\textsuperscript{\textregistered} for a range of parameters showed that in most cases the model is accurate to within 4\% of the actual probability \cite{Mehr}. In summary, the developed model can estimate the probability of reaching a goal state using one or multiple lane changes based on several traffic- and driver-related parameters.
	
	\subsection{Simulation setup} \label{Section2.2}

	The advance warning system proposed in this paper uses the probability model to advise vehicles on when to change lanes to reach a particular goal state, here a highway off-ramp. Specifically, when the vehicle is approaching an off-ramp that it has to take, the system uses traffic data to continuously calculate the probability of reaching that off-ramp and instructs the vehicle to change lanes when the probability dips below a certain threshold. Our goal is to show that if an adequate portion of vehicles use this system, it can help them change lanes on time and reduce overall traffic delay. \par
	
	We used traffic simulations in VISSIM\textsuperscript{\textregistered} to evaluate the performance of the proposed system in reducing delay at a diverge of a four-lane highway. Simulations were carried out for an array of traffic conditions, obtained by varying input traffic flow and the portion of vehicles taking the off-ramp. For each case, we studied how different threshold levels for the probability model (the value at which it advises the driver to start changing lanes) affect overall performance. Details of the traffic simulation setup are presented in \autoref{Section2.2.1} to \autoref{Section2.2.4}.
	
	\subsubsection{Simulation fundamentals} \label{Section2.2.1}
	
\begin{figure}[t!]
		\centering
		\includegraphics[width = \columnwidth]{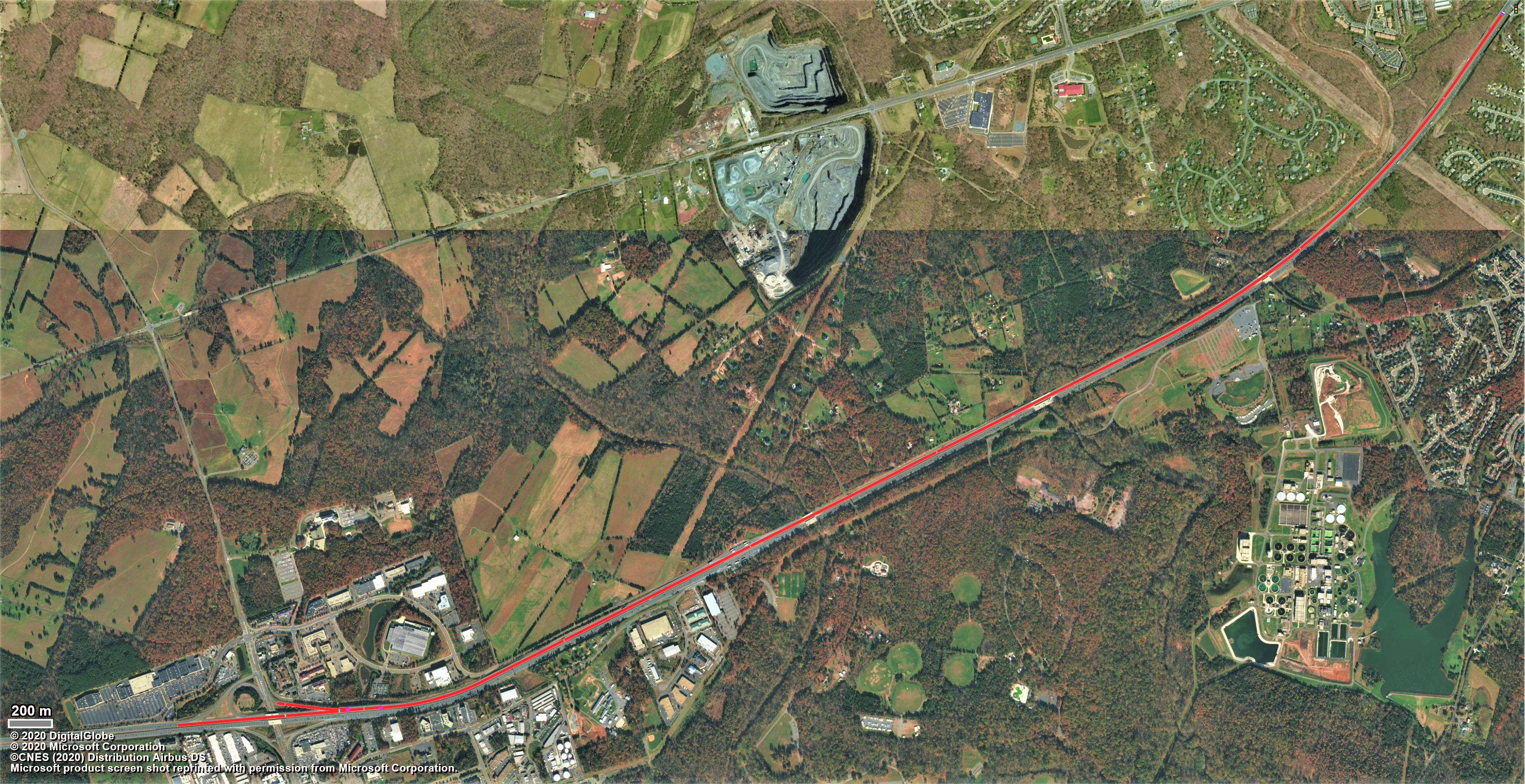}
		\caption{Bird's-eye view of the I-66 highway segment used for traffic simulations. This segment of the westbound interstate road is marked red, starts at the top right corner and ends at the bottom left corner. It is 7.417 kilometers long and has four lanes.} \label{Figure2}
	\end{figure}

\begin{figure}[t!]
		\centering
		\includegraphics[width = \columnwidth]{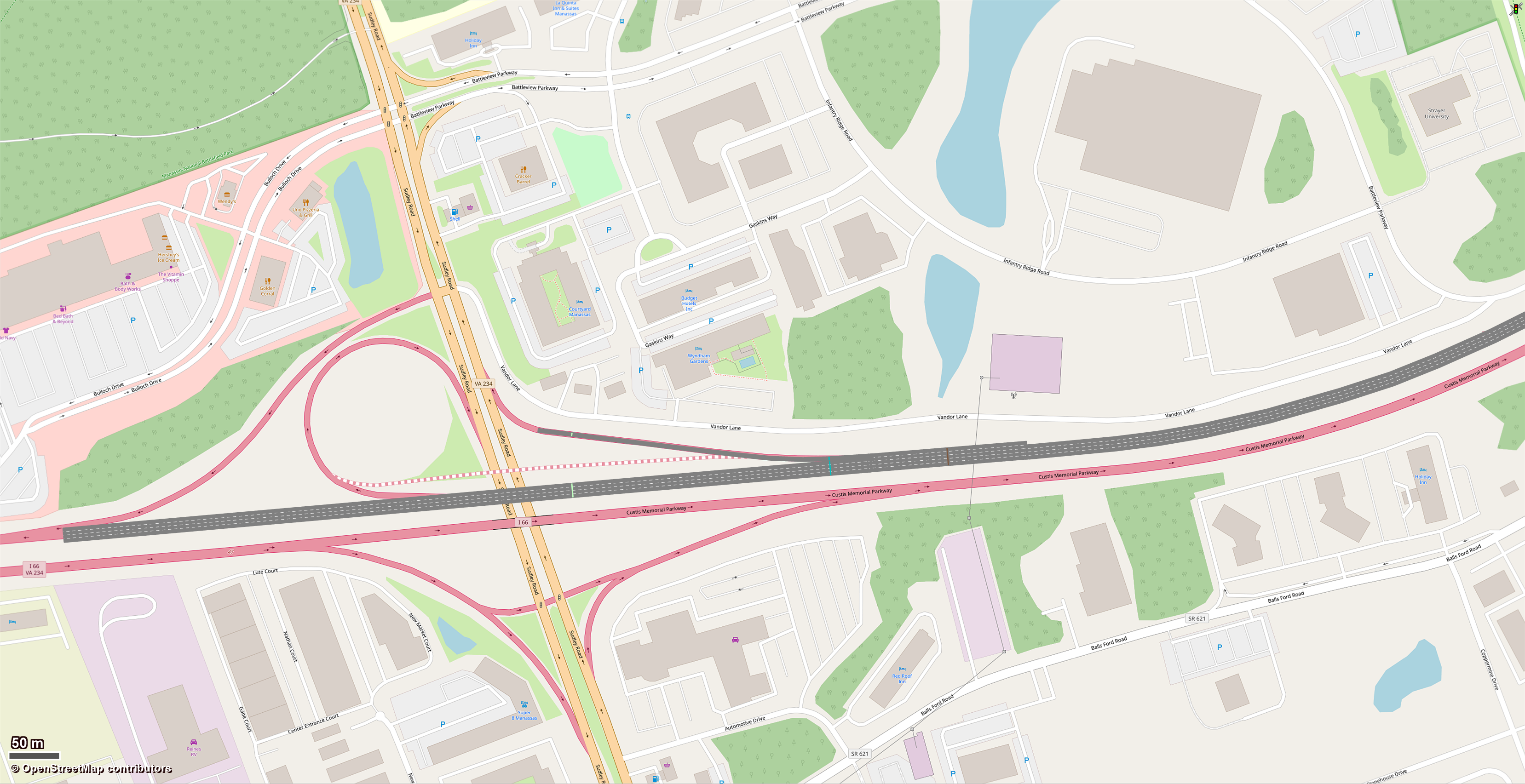}
		\caption{End section of the simulated highway segment. The image shows the deceleration lane leading to the off-ramp to Sudley Road.} \label{Figure3}
	\end{figure}

	Traffic simulations were carried out for a segment of the westbound I-66 interstate highway just outside of Washington D.C., shown in \autoref{Figure2}. It is 7.417 km long and has four lanes. It starts just after the merge from Lee Highway and ends just before the merge from Sudley Road, as shown in \autoref{Figure3}. It has one vehicle input and two vehicle outputs. The input is at the start of the segment and one of the outputs is at the end, while the other one is located at the off-ramp leading to Sudley Road. The deceleration lane to that off-ramp starts at 6.444 km and is 157 meters long. The posted speed limit along the segment is 60 mph (roughly 96.6 km/h), though actual speeds vary based on traffic. \par
	
	
	Input traffic flow, denoted here as $q_{i}$, was set to either 6400 or 9600 vehicles per hour (veh/hr) throughout the simulations, corresponding to a traffic density of 1600 or 2400 vehicles per hour per lane (veh/hr/ln). The volume was stochastic in nature and consisted of three different vehicle types: cars, smart cars, and heavy goods vehicles (HGVs). Smart cars were identical to cars, with the exception that their lane change behavior was controlled by an external driver model (EDM) that advised them on when to change lanes, modeling vehicles that use the proposed advance warning system. For each $q_{i}$, the composition of input vehicles was set as follows: 2\% HGVs, $r$\% smart cars where $r$ was either 2, 6, or 10, and the rest simply cars, resulting in six overall cases. Due to time and computing constraints, $r$ could not be set to a higher value. In all cases both cars and smart cars had a desired speed distribution of 120 km/h at the input, while the desired speed distribution of HGVs was set to 100 km/h. All smart cars were programmed to take the off-ramp, while all other cars continued along the highway. \par
	
	To record traffic data during the simulation, sets of four data collection points were defined at 1 km intervals starting from the 0.5 km mark and ending at the 5.5 km mark. A final set of five data collection points was defined at the middle of the five-lane road segment before the off-ramp consisting of the deceleration lane. These data collection points recorded the time and velocity of vehicles passing through them. Furthermore, a set of travel time measurements was defined to measure vehicle travel times during the simulation. The measurement started at the beginning of the road segment and had two endings, one at the off-ramp and the other at the end of the highway, both a total distance of 6.904 km from the starting point. \par
	
	For each case and each probability threshold $p_{l}$, simulations were run three times, starting from a random seed of 32 and an increment of 5 for successive simulations. Each run lasted 3600 simulation seconds.
	
	\subsubsection{Driving behavior} \label{Section2.2.2}
	
	Driving behavior was defined according to Virginia Department of Transportation's (VDOT) Traffic Operations and Safety Analysis Manual (TOSAM) and VISSIM\textsuperscript{\textregistered} User Guide \cite{Tosam, VissimGuide}. Per the recommendations of \cite{VissimGuide}, the default freeway (free lane selection) driving behavior was used throughout the simulations with parameters shown in \autoref{Table1}. Based on the Wiedemann 99 car-following model, the default parameter values are recommended for freeway segments where significant weaving or merging does not occur \cite{Wiedemann, Fairfax}. 
	
	\begin{table}[b!]
		\caption{Freeway driving behavior parameters} \label{Table1}
		\centering
		\resizebox{0.6\columnwidth}{!}{
		\begin{tabular}{c c}
			\hline
			Parameter & Value \\
			\hline
			CC0 (Standstill Distance) (m) & \phantom{-}1.50 \\
			CC1 (Headway Time) (s) & \phantom{-}0.9\phantom{0} \\
			CC2 (Following Variation) (m) & \phantom{-}4.00 \\
			\hline
			Maximum Deceleration (Own Vehicle) (m/$\text{s} ^ {2}$) & -4.00 \\
			Maximum Deceleration (Trailing Vehicle) (m/$\text{s} ^ {2}$) & -3.00 \\
			Accepted Deceleration (Own Vehicle) (m/$\text{s} ^ {2}$) & -1.00 \\
			Accepted Deceleration (Trailing Vehicle) (m/$\text{s} ^ {2}$) & -0.50 \\
			Safety Distance Reduction Factor & \phantom{-}0.60 \\
			Maximum Deceleration for Cooperative Braking (m/$\text{s} ^ {2}$) & -3.00 \\
			\hline
			Advanced Merging & On \\
			Cooperative Lane Change & Off \\
			\hline
		\end{tabular}}
	\end{table}
	
	The value of Lane Change Distance (distance from a connector that vehicles anticipating a lane change start to act) for the connector leading to the off-ramp was increased from the default value of 200 m to 1600 m. While the 200 m value is usually enough for urban and arterial traffic simulations, it's value should be increased for highway modeling because a small value would result in artificial queues at the diverge \cite{VissimGuide, Gomez}. In the absence of experimental trajectory data to calibrate the model, the 1600 m value (corresponding to the second exit sign on the road) provides a good balance between preventing artificial queues at the diverge and forcing all exiting vehicles to the rightmost lane much earlier than they are supposed to.
	
	\subsubsection{External driver model} \label{Section2.2.3}
	
	VISSIM\textsuperscript{\textregistered}'s External Driver Model (EDM) API grants control over various driving behavior aspects of all or a group of vehicles. For this study, the EDM was used to simulate the proposed on-board advance warning system for an upcoming diverge. \par
	
	For each case (combination of $q_{i}$ and $r$), we first ran the simulation with all vehicles using VISSIM\textsuperscript{\textregistered}'s internal driving behavior, serving as a baseline for later comparison. Using data from data collection points defined earlier, we calculated average values of parameters $v_{i}, \mu_{i}$, and $\sigma_{i}$, $2 \le i \le 4$ as appropriate (depending on the lane the ego vehicle was on), for different road segments. For example, data from the first data collection point (defined at 0.5 km) was used to calculate $v_{i}, \mu_{i}$, and $\sigma_{i}$ for the first 1 km of the road, and so on. As for $g_{i}$, it was set to $\delta v_{i} + s_{0}$ where for our simulations $\delta$ and $s_{0}$ were set to 1.6 s and 1 m, respectively. Though in reality the critical gap used by drivers is stochastic in nature and depends on a variety of factors - including relative speeds of leading and trailing vehicles in the adjacent lane and driver aggressiveness - our choice simplifies the model and its conservative nature (generally being larger than the actual critical gap) makes sure unsafe lane changes do not occur \cite{Toledo}. Finally, $t_{i}$ was set to 3 seconds, as VISSIM\textsuperscript{\textregistered}'s internal model completes a lane change in that time from when it is initiated \cite{PTV}. \par
	
	In subsequent simulations for each case, smart cars used the EDM to advise them on when to change lanes. Along with the values of $v_{i}, \mu_{i}, \sigma_{i}, g_{i}$, and $t_{i}$, for each vehicle the EDM used its distance to the off-ramp as $d$ and its velocity as $v_{1}$. The only exception to this process was when $v_{i + 1}$ was within the closed interval defined by endpoints $v_{i} \pm v_{l}$, where $v_{l}$ was set to 4 m/s. In that case, $v_{i + 1} = v_{i} + v_{l}$. This was done because our previous work in \cite{Mehr} showed that when $v_{i}$ and $v_{i + 1}$, $1 \le i \le n - 1$, are close to each other, the probability drops significantly due to a large reduction in the relative traveled distance which is unrealistic. Therefore, this modification was made to more accurately represent the accelerating or decelerating behavior of drivers when looking for a gap in an adjacent lane during a lane change. \par
	
	To simulate the proposed advance warning system, the EDM was programmed to continuously calculate the probability of reaching the off-ramp for each smart car. If the probability dipped below a certain threshold $p_{l}$, the EDM instructed that vehicle to change lanes. For each case, we tested $p_{l}$ values of 0.99, 0.95, 0.9, 0.85, 0.8, and 0.75 to understand its effect on traffic flow and average delay. \par
	
	One problem that we faced was that whenever the EDM instructed a vehicle to change lanes, VISSIM\textsuperscript{\textregistered} would immediately start to do so without first checking if it was safe. To solve this problem, the EDM first checked to see if conducting a lane change was safe before instructing a vehicle to do so. It used the velocity of the ego vehicle relative to its leading and trailing vehicles in the adjacent lane to calculate the leading and trailing critical gaps given in (\ref{Equation2}) and (\ref{Equation3}) and compared them to the relative distance between the ego vehicle and its leading and trailing vehicles in the adjacent lane \cite{Toledo}. If both distances were larger than the critical gap, it would proceed with the lane change.
	\begin{align}
		g_{i} ^ {\mathrm{lead, cr}} &= \exp\big(1.353 - 2.700\max[0, \Delta v_{i} ^ {\mathrm{lead}}] - 0.231\min[0, \Delta v_{i} ^ {\mathrm{lead}}] + \epsilon ^ {\mathrm{lead}}\big), \label{Equation2} \\
		g_{i} ^ {\mathrm{lag, cr}} &= \exp\big(1.429 + 0.471\max[0, \Delta v_{i} ^ {\mathrm{lag}}] + \epsilon ^ {\mathrm{lag}}\big), \label{Equation3}
	\end{align}
	where $\epsilon ^ {\mathrm{lead}}\sim N(0, 1.112 ^ {2})$ and $\epsilon ^ {\mathrm{lag}}\sim N(0, 0.742 ^ {2})$. In the equations above, $g ^ {\mathrm{lead}}$ refers to the gap between the front bumper of the ego vehicle and the rear bumper of the leading vehicle in the adjacent lane and $g ^ {\mathrm{lag}}$ refers to the gap between the rear bumper of the ego vehicle and the front bumper of the trailing vehicle in the adjacent lane. Similarly, $\Delta v ^ {\mathrm{lead}}$ and $\Delta v ^ {\mathrm{lag}}$ refer to the velocity of the leading and trailing vehicles in the adjacent lane relative to the velocity of the ego vehicle, respectively. Finally, $\epsilon$ is a random term associated with lane utility \cite{Toledo}.
	
	\subsubsection{Data processing and evaluation} \label{Section2.2.4}
	
	Average delay, defined as the difference between actual travel time and travel time at free flow speed, was selected as our measure of effectiveness (MoE) \cite{VissimGuide}. Using the travel time measurement defined in \autoref{Section2.2.1}, VISSIM\textsuperscript{\textregistered} automatically calculated the delay for each individual vehicle. Using that data, we calculated the average, standard deviation and maximum delay for each run, reporting the three-run average for each combination of $q_{i}, r$, and $p_{l}$.
	
	\section{Results and Discussion} \label{Section3}
	
	\renewcommand{\tabcolsep}{4 pt}
	\begin{table}[b!]
		\caption{Statistical characteristics of traffic delay results for all vehicles (numbers in parenthesis show percentage change relative to the baseline case).} \label{Table2}
		\resizebox{\columnwidth}{!}{
		\begin{tabular}{c c c c c c c c c c c}
			\hline
			\multirow{2}{*}{$q_{i}$ (veh/hr)} & \multirow{2}{*}{$p_{l}$} & \multicolumn{3}{c}{$r = 2$} & \multicolumn{3}{c}{$r = 6$} & \multicolumn{3}{c}{$r = 10$} \\
			\cline{3 - 11}
			 &  & Avg. (s) & Std. (s) & Max. (s) & Avg. (s) & Std. (s) & Max. (s) & Avg. (s) & Std. (s) & Max. (s) \\
			\hline
			\multirow{7}{*}{6400} & baseline & 12.94 & 14.10 & 101.00 & 12.96 & 14.09 & 104.50 & 13.10 & 14.14 & 103.10 \\
			 & 0.99 & 12.93 (-0.04) & 14.05 (-0.35) & 96.57 (-4.39) & 12.92 (-0.31) & 14.00 (-0.67) & 101.57 (-2.81) & 13.00 (-0.69) & 13.95 (-1.31) & 86.97 (-15.65) \\
			 & 0.95 & 12.93 (-0.07) & 14.06 (-0.28) & 96.57 (-4.39) & 12.88 (-0.60) & 13.96 (-0.95) & 98.63 (-5.61) & 12.98 (-0.90) & 13.94 (-1.37) & 93.63 (-9.18) \\
			 & 0.9 & 12.90 (-0.29) & 14.05 (-0.35) & 96.70 (-4.26) & 12.86 (-0.71) & 13.93 (-1.12) & 97.77 (-6.44) & 12.95 (-1.12) & 13.93 (-1.44) & 90.57 (-12.16) \\
			 & 0.85 & 12.91 (-0.23) & 14.05 (-0.33) & 96.20 (-4.75) & 12.87 (-0.67) & 13.94 (-1.07) & 97.77 (-6.44) & 12.94 (-1.18) & 14.57 (3.3) & 96.93 (-5.98) \\
			 & 0.8 & 12.90 (-0.24) & 14.05 (-0.30) & 96.23 (-4.72) & 12.86 (-0.74) & 13.94 (-1.09) & 98.07 (-6.16) & 12.94 (-1.18) & 13.95 (-1.33) & 96.53 (-6.37) \\
			 & 0.75 & 12.91 (-0.17) & 14.06 (-0.29) & 96.20 (-4.75) & 12.85 (-0.78) & 13.93 (-1.13) & 99.23 (-5.04) & 12.93 (-1.27) & 13.94 (-1.38) & 99.23 (-3.75) \\
			\hline
			\multirow{7}{*}{9600} & baseline & 28.78 & 19.63 & 112.37 & 28.37 & 19.44 & 113.67 & 29.51 & 19.78 & 111.67 \\
			 & 0.99 & 28.62 (-0.56) & 19.79 (0.78) & 109.50 (-2.55) & 29.11 (2.57) & 19.51 (0.34) & 109.97 (-3.26) & 30.00 (1.66) & 19.84 (0.30) & 113.93 (2.03) \\
			 & 0.95 & 27.68 (-3.82) & 19.24 (-1.99) & 109.63 (-2.43) & 28.43 (0.21) & 19.40 (-0.19) & 103.37 (-9.06) & 28.82 (-2.33) & 19.32 (-2.35) & 104.73 (-6.21) \\
			 & 0.9 & 28.92 (0.51) & 19.59 (-0.20) & 115.87 (3.11) & 28.61 (0.83) & 19.32 (-0.62) & 112.70 (-0.85) & 27.76 (-5.92) & 19.14 (-3.25) & 108.37 (-2.96) \\
			 & 0.85 & 29.15 (1.29) & 19.64 (0.03) & 114.20 (1.63) & 27.96 (-1.46) & 19.25 (-0.97) & 111.97 (-1.50) & 27.86 (-5.58) & 19.11 (-3.40) & 108.87 (-2.51) \\
			 & 0.8 & 29.01 (0.81) & 19.52 (-0.57) & 108.17 (-3.74) & 28.14 (-0.82) & 19.32 (-0.63) & 112.17 (-1.32) & 27.66 (-6.28) & 19.04 (-3.75) & 107.23 (-3.97) \\
			 & 0.75 & 29.34 (1.96) & 19.58 (-0.29) & 112.23 (-0.12) & 28.35 (-0.09) & 19.33 (-0.59) & 111.40 (-1.99) & 28.29 (-4.12) & 19.19 (-2.99) & 106.03 (-5.04) \\
			\hline
		\end{tabular}}
	\end{table}
	\renewcommand{\tabcolsep}{4 pt}
	\begin{table}[t!]
		\caption{Statistical characteristics of traffic delay results for smart cars (numbers in parenthesis show percentage change relative to the baseline case).} \label{Table3}
		\resizebox{\columnwidth}{!}{
		\begin{tabular}{c c c c c c c c c c c}
			\hline
			\multirow{2}{*}{$q_{i}$ (veh/hr)} & \multirow{2}{*}{$p_{l}$} & \multicolumn{3}{c}{$r = 2$} & \multicolumn{3}{c}{$r = 6$} & \multicolumn{3}{c}{$r = 10$} \\
			\cline{3 - 11}
			 &  & Avg. (s) & Std. (s) & Max. (s) & Avg. (s) & Std. (s) & Max. (s) & Avg. (s) & Std. (s) & Max. (s) \\
			\hline
			\multirow{7}{*}{6400} & baseline & 15.45 & 15.44 & 67.47 & 15.48 & 15.42 & 76.43 & 15.85 & 15.41 & 76.50 \\
			 & 0.99 & 16.37 (5.92) & 15.83 (2.53) & 70.20 (4.05) & 16.80 (8.48) & 15.86 (2.83) & 70.80 (-7.37) & 17.01 (7.34) & 15.73 (2.07) & 69.27 (-9.46) \\
			 & 0.95 & 15.78 (2.14) & 15.61 (1.08) & 66.47 (-1.48) & 15.67 (1.24) & 15.28 (-0.94) & 65.50 (-14.30) & 16.37 (3.27) & 15.47 (0.37) & 69.17 (-9.59) \\
			 & 0.9 & 15.36 (-0.57) & 15.52 (0.53) & 66.03 (-2.12) & 15.39 (-0.57) & 15.18 (-1.57) & 64.87 (-15.13) & 15.84 (-0.05) & 15.24 (-1.12) & 68.83 (-10.02) \\
			 & 0.85 & 15.32 (-0.88) & 15.48 (0.24) & 66.07 (-2.08) & 15.23 (-1.64) & 15.08 (-2.20) & 65.93 (-13.74) & 15.76 (-0.58) & 15.19 (-1.42) & 69.33 (-9.37) \\
			 & 0.8 & 15.30 (-1.01) & 15.47 (0.19) & 65.67 (-2.67) & 15.12 (-2.37) & 15.02 (-2.61) & 65.67 (-14.09) & 15.67 (-1.12) & 15.13 (-1.82) & 69.17 (-9.59) \\
			 & 0.75 & 15.27 (-1.17) & 15.42 (-0.13) & 65.67 (-2.67) & 15.10 (-2.44) & 15.02 (-2.64) & 64.97 (-15.00) & 15.56 (-1.82) & 15.09 (-2.09) & 69.10 (-9.67) \\
			\hline
			\multirow{7}{*}{9600} & baseline & 29.80 & 19.51 & 84.00 & 29.92 & 19.72 & 97.53 & 32.25 & 19.96 & 102.60 \\
			 & 0.99 & 32.84 (10.20) & 20.61 (5.67) & 91.73 (9.21) & 34.80 (16.31) & 21.03 (6.67) & 105.97 (8.65) & 37.85 (17.37) & 21.51 (7.74) & 109.97 (7.18) \\
			 & 0.95 & 30.28 (1.63) & 19.50 (-0.03) & 86.27 (2.70) & 32.06 (7.16) & 20.24 (2.66) & 95.50 (-2.08) & 33.78 (4.76) & 20.09 (0.67) & 100.53 (-2.01) \\
			 & 0.9 & 30.38 (1.97) & 19.56 (0.30) & 88.30 (5.12) & 31.39 (4.90) & 19.94 (1.11) & 96.70 (-0.85) & 31.69 (-1.73) & 19.64 (-1.59) & 98.60 (-3.90) \\
			 & 0.85 & 30.31 (1.72) & 19.56 (0.29) & 89.57 (6.63) & 30.21 (0.96) & 19.68 (-0.17) & 92.23 (-5.43) & 31.44 (-2.49) & 19.40 (-2.80) & 98.23 (-4.26) \\
			 & 0.8 & 29.52 (-0.94) & 19.11 (-2.04) & 86.63 (3.13) & 30.55 (2.09) & 19.85 (0.66) & 98.57 (1.06) & 31.21 (-3.23) & 19.40 (-2.80) & 98.23 (-4.26) \\
			 & 0.75 & 30.02 (0.73) & 19.38 (-0.63) & 86.17 (2.58) & 30.25 (1.11) & 19.81 (0.48) & 96.90 (-0.65) & 31.67 (-1.80) & 19.51 (-2.25) & 97.83 (-4.65) \\
			\hline
		\end{tabular}}
	\end{table}
	
	Statistical characteristics of traffic delay results are tabulated in \autoref{Table2} for all vehicles and in \autoref{Table3} for smart cars as a subset of all vehicles. In each table the results are divided according to $q_{i}$ in the first column and $r$ in the first row. For each combination of these two parameters, the second column shows different values used for $p_{l}$, with baseline being the case where no advance warning system is present. In each block defined by the combination of parameters $q_{i}$ and $r$, the three columns present the average, standard deviation of, and maximum delay. For each row other than baseline, the numbers in parenthesis show percentage change relative to the baseline case. For example, in \autoref{Table2} the third, fourth, and fifth columns of the fifth row show the average, standard deviation of, and maximum delay for the simulation case with $q_{i}$ = 6400 veh/hr, $r$ = 2, and $p_{l}$ = 0.95, with the numbers in parenthesis showing percentage change relative to the baseline case in the third row. For this example, average and maximum delay were improved by 0.07\% and 4.39\% relative to the baseline case, respectively. \par
	
	An overall look at the results shows that in most cases the system is successful at reducing traffic delay regardless of the probability threshold, albeit in some cases not by a large margin. When the $q_{i}$ is smaller and vehicles are more spread out, as is the case for $q_{i}$ = 6400 veh/hr, changing lanes causes less delay, so average traffic delay (and its standard deviation) is relatively small and any improvements would be small and within the margin of error. On the other hand, when the road is at near full capacity (about 2400 veh/hr/ln), lane-change-induced delays increase and there is more room for improvement, as seen in the results. \par
	
	Focusing on $r$, a trend that emerges is that as the portion of cars taking the off-ramp increase, the average delay of the baseline case (and its standard deviation) slightly increases, but so too does the improvement in it when using an advance warning system. For example, when $q_{i}$ = 6400 veh/hr, for $r$ = 2\% the improvement in average delay is at most 0.29\% which is negligible, but for $r$ = 10\% the improvement can reach up to 1.27\%. This makes sense, because when a larger portion of the vehicles are advised by the advance warning system and plan ahead, improvements in traffic delay will be larger. As mentioned previously, we were not able to simulate cases with a larger $r$ due to computing and time constraints, but it would be interesting to see if this trend holds for those larger values, for example when $r = 30$.
	
	\begin{figure}[b!]
		\centering
			\begin{tikzpicture}
				\begin{axis}[title = {}, xlabel = {Distance (km)}, ylabel = {$N$}, xmin = 0, xmax = 6000, ymin = 0, ymax = 500, xtick = {0, 1000, 2000, 3000, 4000, 5000, 6000}, xticklabels = {0, 1, 2, 3, 4, 5, 6}, ytick = {0, 100, 200, 300, 400, 500}, xlabel near ticks, ylabel near ticks, legend pos = north west, legend style = {font = \footnotesize}, xmajorgrids = true, ymajorgrids = true, grid style = dashed, width = 0.6\columnwidth]
					\addplot[hist = {bins = 120, data min = 0, data max = 6000}, color = red, fill = red, fill opacity = 0.2, no marks] table[y = LP1] {Data/Figure4.txt};
					\addplot[hist = {bins = 120, data min = 0, data max = 6000}, color = blue, fill = blue, fill opacity = 0.2, no marks] table[y = LP2] {Data/Figure4.txt};
					\legend{$p_{l}$ = 0.99, $p_{l}$ = 0.8};
				\end{axis}
			\end{tikzpicture}
		\caption{Histogram of the location of the last lane change of smart cars before reaching the rightmost lane for two $p_{l}$ values for the case with $q_{i}$ = 9600 veh/hr and $r$ = 10.} \label{Figure4}
	\end{figure}
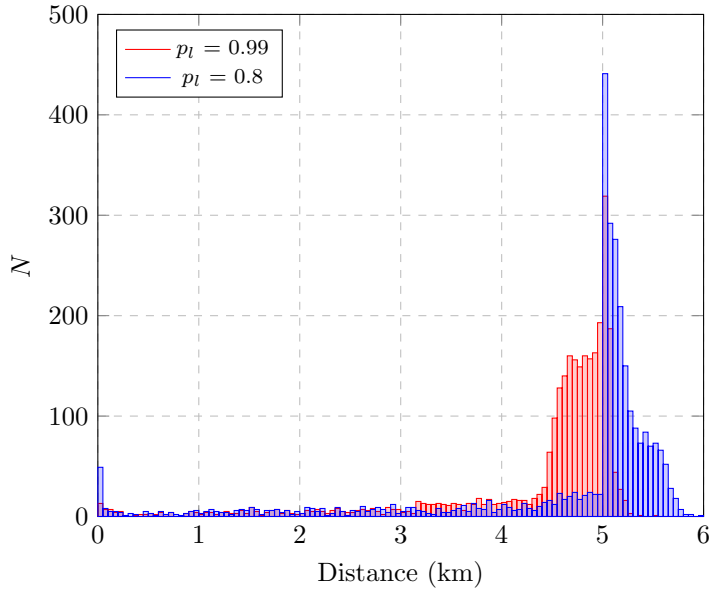
	
	At first glance, results in \autoref{Table2} do not reveal a discernible relationship between $p_{l}$ and the amount of improvement in average delay and its standard deviation. To explain this, we need to look at \autoref{Table3}. For each baseline case, average delay for smart cars is slightly higher than the overall average delay, which is expected since these vehicles are likely to change lanes to get to the off-ramp, hence encountering delay. However, when looking at the results of non-baseline iterations, one can see that average delay decreases as $p_{l}$ decreases (the overall relationship between $p_{l}$ and average delay seems to be parabolic, especially when $q_{i}$ is large. Look, for example, at the results in \autoref{Table3} for $q_{i}$ = 9600 veh/hr and $r$ = 10). To explain this, one should note that a higher $p_{l}$, for example 0.99, means that smart cars generally change lanes sooner than when $p_{l}$ is lower, say 0.8. This can be seen in \autoref{Figure4}, which shows a histogram of the location of the last lane change of smart cars before reaching the rightmost lane for the case with $q_{i}$ = 9600 veh/hr and $r$ = 10. When $p_{l}$ = 0.99, most lane changes occur before the 5 km mark, but when $p_{l}$ = 0.8 most lane changes occur afterwards. Therefore, when $p_{l}$ is high smart cars may have to drive in the rightmost lane - which may be slower - for a longer period, increasing their average delay at the expense of the delay of all other vehicles. On the other hand, when $p_{l}$ is low and exiting vehicles make lane changes later, perhaps in a rushed way, their average delay is decreased at the expense of all other vehicles that need to slow down for the lane change to take place. For this reason, the best overall probability threshold is one that provides a balance between the delay of smart cars and those of all other cars. \par
	
	Finally, \autoref{Table2} reveals that the advance warning system has a consistently larger impact on reducing maximum delay compared to average delay for each case. Large delays are generally caused by those vehicles that change lanes very late, having to decelerate and slow down vehicles around them significantly. By using the advance warning system and being aware of when to change lanes to reach the off-ramp on time, exiting vehicles can avoid those situations, reducing the maximum delay. 
	
	\section{Conclusions and Outlook} \label{Section4}
	
	This paper introduced an advance warning system for vehicles based on a probabilistic prediction model that advises them on when to change lanes to reach a highway diverge on time. After briefly discussing the probability model, we laid out the VISSIM\textsuperscript{\textregistered} simulation setup used to understand the effects of employing this system on traffic delay at a 4-lane highway diverge. Vigorous simulations for cases with different traffic flows and vehicle compositions showed that the system was successful in reducing average and maximum traffic delay regardless of the probability threshold, though the amount of improvement depended on it. In the real world, this system can be a simple addition to navigation software that use traffic data to advise vehicles on a highway to change lanes to reach their goals, improving traffic flow and reducing delay in the process. \par
	
	As current simulation results have been promising, future work will focus on studying the impact of this system on driving behavior using full-cabin driving simulators. Furthermore, recognizing that the underlying probability model is not limited to the case of highway diverges, our future work focuses on applying it as an advisory system for upcoming road features such as lane drops, and incidents such as work zones and traffic accidents.
	
	\section*{Acknowledgment} \label{Section5}
	
	The authors wish to express their gratitude to Dr. Harpreet S. Dhillon for his help with the probability model and to Dr. Montasir Abbas and Awad Abdelhalim for their help with VISSIM\textsuperscript{\textregistered} simulations.
	
	\bibliographystyle{iet}
	\bibliography{BIB}

\end{document}